\documentclass[useAMS,usenatbib]{mnras}

\usepackage{newtxtext,newtxmath}

\usepackage[T1]{fontenc}
\usepackage{ae,aecompl}

\usepackage{graphicx}	
\usepackage{multicol}        
\usepackage{bm}		
\usepackage{pdflscape}	

\DeclareMathAlphabet{\mathsc}{OT1}{cmr}{m}{sc}
\def\testbx{bx}%
\DeclareRobustCommand{\ion}[2]{%
\relax\ifmmode
\ifx\testbx\f@series
{\mathbf{#1\,\mathsc{#2}}}\else
{\mathrm{#1\,\mathsc{#2}}}\fi
\else\textup{#1\,{\mdseries\textsc{#2}}}%
\fi}

\title[Surviving companion stars of SNe Iax]{Long-term evolution of surviving companion stars of Type I\lowercase{ax} supernovae}

\author[Z.-W. Liu \& Y. Zeng]{
Zheng-Wei Liu$^{1,2,3,4}$\thanks{E-mail: zwliu@ynao.ac.cn} and Yaotian Zeng$^{1,2,4}$
\\
$^{1}$Yunnan Observatories, Chinese Academy of Sciences (CAS), 396 Yangfangwang, Guandu District, Kunming 650216, P.R. China\\
$^{2}$Key Laboratory for the Structure and Evolution of Celestial Objects, CAS, Kunming 650216, P.R. China\\
$^{3}$Center for Astronomical Mega-Science, CAS, Beijing 100012, P. R. China\\
$^{4}$University of Chinese Academy of Science, Beijing 100012, P.R. China\\
}

\pubyear{2020}

\begin{document}
\label{firstpage}
\pagerange{\pageref{firstpage}--\pageref{lastpage}}
\maketitle

\begin{abstract}

The nature of the progenitors and explosion mechanism of Type Iax supernovae (SNe Iax) remain a mystery. The single-degenerate (SD) systems that involve the incomplete pure deflagration explosions of near-Chandrasekhar-mass white dwarfs (WDs) have recently been proposed for producing SNe Iax, in which non-degenerate companions are expected to survive from SN explosions. In this work we concentrate on the main-sequence (MS) donor SD progenitor systems. By mapping the computed companion models from three-dimensional hydrodynamical simulations of ejecta-companion interaction into a one-dimensional stellar evolution code \textsc{MESA}, we investigate the long-term appearance and observational signatures of surviving MS companions of SNe Iax by tracing their post-impact evolution. Depending on different MS companion models, it is found that the shocked surviving companion stars can significantly expand and evolve to be more luminous ($5$--$500\,L_{\sun}$) for a time-scale of $10$--$10^{4}$ yr. Comparing with the late-time light curve of an observed SN Iax (SN~2005hk), it is suggested that surviving MS companions of SNe Iax would expect to be visible about 1000~d after the explosion when SN itself has been faded.

\end{abstract}




\section{Introduction} \label{sec:introduction}

Type Ia supernovae (SNe~Ia) are thought to be resulted from thermonuclear explosions of white dwarfs (WDs) in binary systems \citep{Hoyle1960}. Although SNe Ia play a fundamental role in astrophysics, the nature of their progenitor systems and the explosion mechanism remain an unsolved mystery \citep[e.g.][]{Hillebrandt2000, Maoz2014, Livio2018}. To explain observational features of SNe Ia, a set of progenitor models which cover different explosion mechanism have been proposed for SNe Ia, including the single-degenerate (SD; \citealt{Whelan1973, Han2004}), the double-degenerate (DD; \citealt{Iben1984, Webbink1984, Dan2011, Dan2012}), the sub-Chandrasekhar mass \citep[e.g.][]{Nomoto1982, Woosley1986, Fink2007, Shen2007, Sim2010, Gronow2020}, the violent merger \citep{Pakmor2010, Pakmor2012b}, the core-degenerate \citep[e.g.][]{Livio2003,Kashi2011,Ilkov2012}, and the WD-WD head-on collisions model \citep[e.g.][]{Benz1989, Rosswog2009,Kushnir2013}.

In the SD scenario, the WDs accrete matter from a non-degenerate companion which could be a main-sequence (MS), a subgiant, a red giant, and/or a helium star to trigger thermonuclear explosions when they grow in mass to approach the Chandrasekhar-mass limit \citep{Whelan1973}. The companion stars are expected to survive the explosion in this scenario. Therefore, searching for the surviving companion star in galactic SN remnants (SNRs) has been proposed to be a promising approach to test the progenitors of SNe Ia \citep[e.g.][]{Pan2014, Ruiz-Lapuente2019}. To explore the impact of SN Ia explosion on its companion, the interaction of SN Ia ejecta with a stellar companion star has been studied in detail by the analytical method, two-dimensional (2D) and/or three-dimensional (3D) hydrodynamical simulations \citep[e.g.][]{Wheeler1975, Marietta2000, Pakmor2008, Liu2012, Liu2013a, Liu2013c, Pan2012a,Boehner2017,Bauer2019, Zeng2020}. There are several ways by which the SN blast wave modifies the properties of the companion star. First, some matters are removed from the surface of the companion star and a significant amount of thermal energy is injected into star during the interaction with SN ejecta. This causes that the surviving companion star dramatically puffs up and becomes overluminous. For instance, By tracing long-term evolution of surviving MS companions produced from 3D hydrodynamical impact simulations, \citet[][]{Pan2012b} found that they dramatically expands and become overluminous (up to about $100\,{L_{\sun}}$) within a few thousand years, depending on the progenitor model \citep[see also][]{Podsiadlowski2003, Shappee2012}. Secondly, the companion star's surface could be enriched with heavy elements (e.g. $\mathrm{Ni}$, $\mathrm{Fe}$) which come from the inner part of the SN ejecta \citep[e.g.][]{Gonzalez-Hernandez2009,Liu2012, Liu2013a, Pan2013}. This is expected to present some features of these heavy elements in spectra of a surviving companion star. In addition, the binary system is destroyed while the SN explodes, the companion star releases and moves in a peculiar velocity compared to other stars in the vicinity, becoming a hypervelocity (or runaway) star \citep[e.g.][]{Geier2015}.

\begin{figure*}
  \begin{center}
    {\includegraphics[width=0.96\textwidth, angle=360]{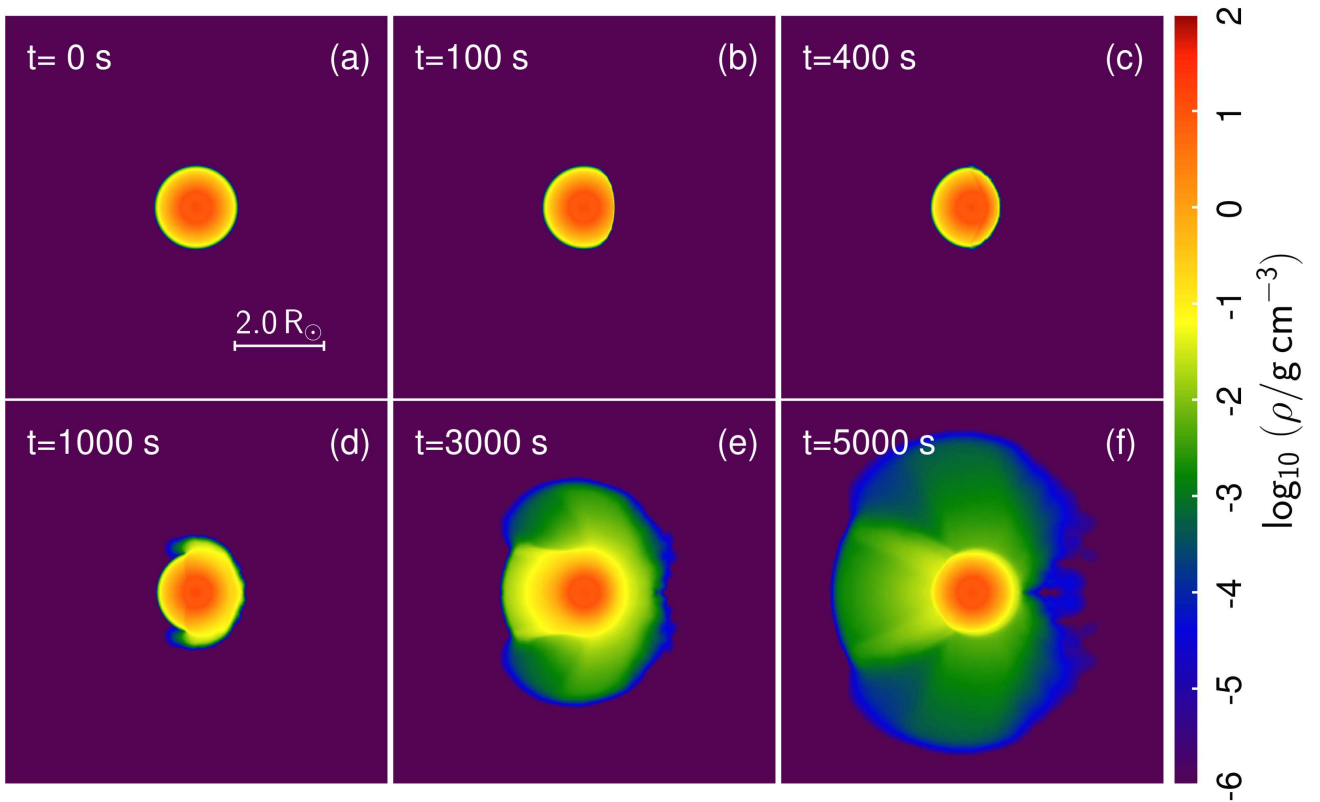}}
    \caption{Density distributions of bound companion material in the x-y plane as a function of time in 3D hydrodynamical simulations of SN~ejecta-companion interaction for the Model\_A \citep[][see their Table~1]{Liu2013c}. The color scale shows the logarithm of the mass density in $\rm{g\,cm^{-3}}$ .}
\label{fig:remnant}
  \end{center}
\end{figure*}

On the observational side, many studies have been focused on searching for the surviving companion star predicted by the SD scenario in SNRs \citep[for a recent review, see][]{Ruiz-Lapuente2019}. But unfortunately, to date, no promising surviving companion star has been identified yet in SNRs  \citep[e.g.][]{Kerzendorf2009, Schaefer2012, Gonzalez-Hernandez2012, Ruiz-Lapuente2019}. Although it has been suggested that Tycho~G is a possible candidate because of some peculiar features such as its lower surface gravity than a MS star, a peculiar velocity in radial and proper motion \citep{Ruiz-Lapuente2004, Bedin2014}, and an overabundance of $\mathrm{Ni}$ relative to normal metal-rich stars \citep{Gonzalez-Hernandez2009}, some other studies casts some doubts on this identification. For instance, \citet{Howell2011} concluded that Tycho~G is apparently not out of thermal equilibrium. In addition, \citet{Kerzendorf2013} showed that the measured $\mathrm{[Ni/Fe]}$ ratio of Tycho~G seems to be not so unusual with respect to field stars with the same metallicity.

Type Iax supernovae (SNe Iax) are proposed as one new subclass since they present sufficiently distinct observational properties from the bulk of SNe Ia \citep[e.g.][]{Li2003, Foley2013}. It has been found that they contribute about 1/3 of total SNe Ia \citep[e.g.][]{Li2001,Foley2013}. SNe Iax are fainter compared to normal SNe Ia and highly skewed to late-type galaxies \citep[e.g.][]{Foley2013,Liu2015b}. Their explosion ejecta are characterized by low expansion velocities and show strong mixing features. In addition, strong He lines are identified in spectra of two events, i.e., SN~2004cs and SN~2007J, and late-time spectra of SNe Iax are dominated by narrow permitted Fe~II \citep{Jha2006}. We refer to \citet{Foley2013} and \citet{Jha2017} for a detailed introduction of observational features of SNe Iax.

Although the progenitors and explosion mechanism of SNe Iax remain a mystery, it has been suggested that incomplete pure deflagration explosions of near-Chandrasekhar-mass WDs within the SD scenario are able to reproduce the characteristic observational features of SNe Iax \citep[e.g.][]{Jordan2012a, Kromer2013, Kromer2015, Fink2014, Liu2015, Bulla2020}. Recently, using the smoothed-particle hydrodynamics (SPH) code \textsc{Stellar Gadget} \citep{Springel2001, Pakmor2012a}, \citet{Liu2013c} have performed 3D hydrodynamical simulations of SN~ejecta-companion interaction by assuming that SNe Iax are generally generated from pure deflagration explosions of near-Chandrasekhar-mass WDs in WD~+~MS progenitor systems. In that work, they concentrated on calculating the amount of stripped H mass from companion surface by SN Iax explosion. They found that H mass stripping in SNe Iax is in inefficient ($\lesssim0.01\,M_{\sun}$) due to the low kinetic energy of the adopted explosion model, i.e., the N5def model of \citet{Kromer2013}\footnote{In the N5def model, only a part of the Chandrasekhar-mass WD is ejected, leading to a ejecta mass of about $0.37\,M_{\sun}$ and a kinetic energy of $1.34\times10^{50}\,\mathrm{erg}$. It has been found that this model is able to reproduce the characteristic observational features of a prototypical 2002cx-like event, SN 2005hk \citep{Kromer2013}}, suggesting that H lines caused by stripped material from a MS companion star may be hidden in late-time spectra of SNe Iax \citep{Tucker2020, Jacobson2019}. However, the long-term evolution of surviving MS companions was not addressed in detail in their work.

\begin{figure}
  \begin{center}
    {\includegraphics[width=0.46\textwidth, angle=360]{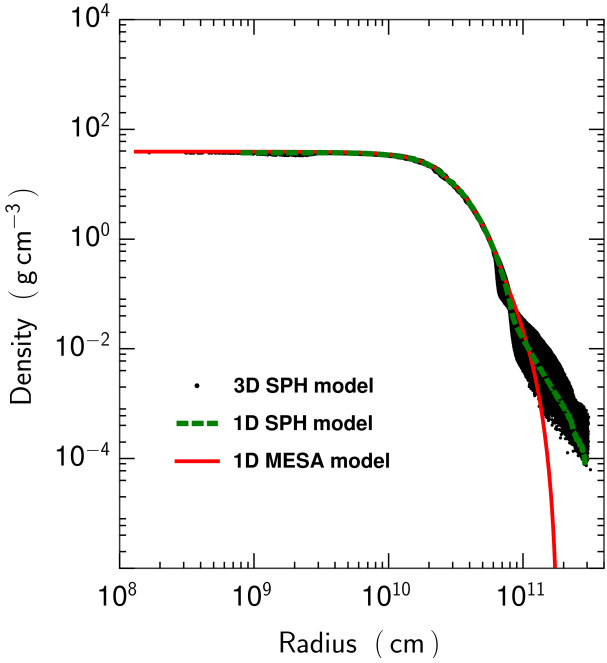}}
    \caption{Density profiles of a surviving companion star for Model\_D of \citet{Liu2013c}. The black dots, green dashed curve, and red solid curve are corresponding to the 3D SPH model, the 1D angle-averaged profile of SPH model before the relaxation, and the relaxed starting model of \textsc{MESA}, respectively.}
\label{Fig:2}
  \end{center}
\end{figure}

\begin{figure*}
  \begin{center}
    {\includegraphics[width=0.48\textwidth, angle=360]{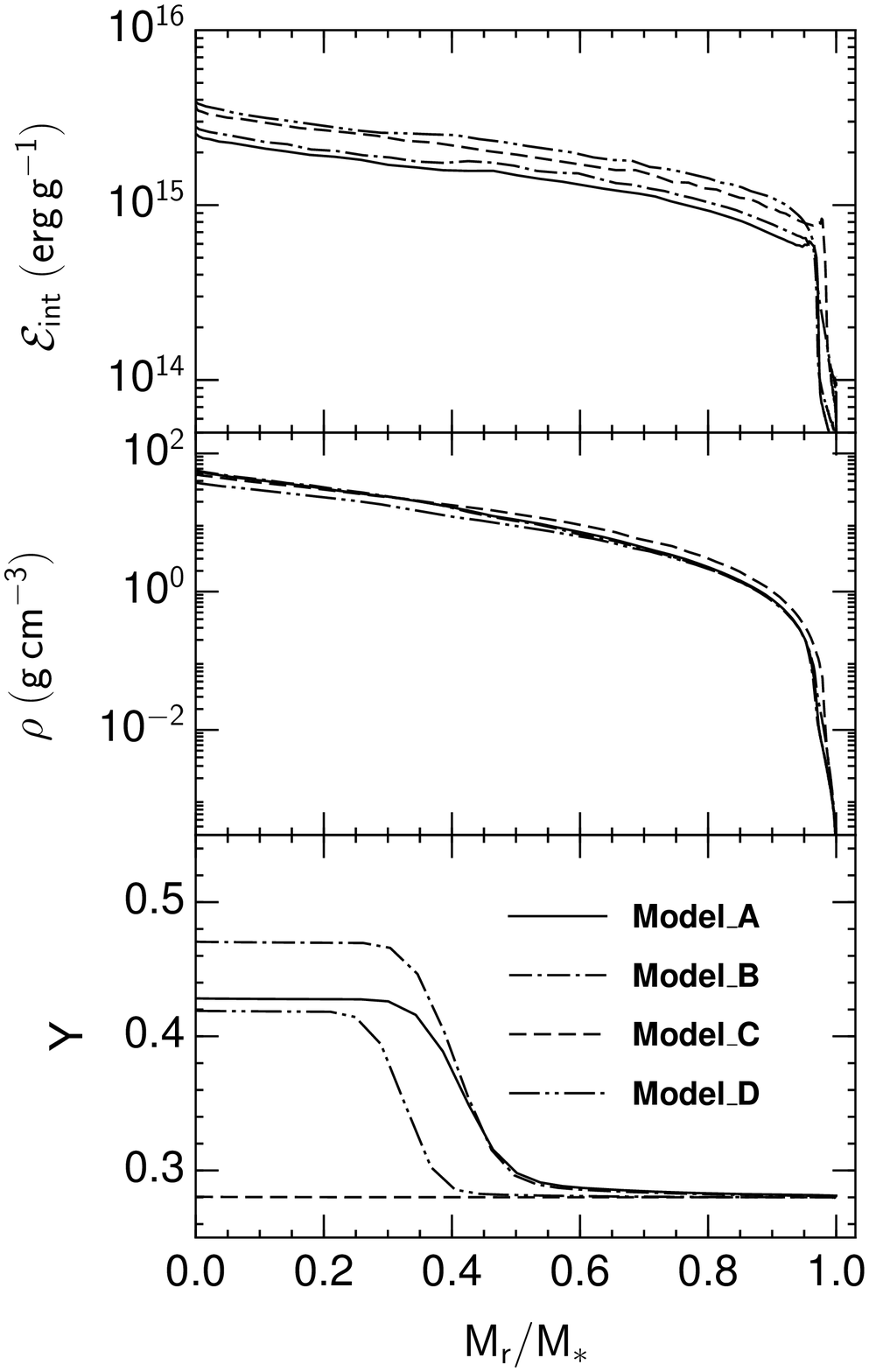}}
    \hspace{0.2in}
   {\includegraphics[width=0.48\textwidth, angle=360]{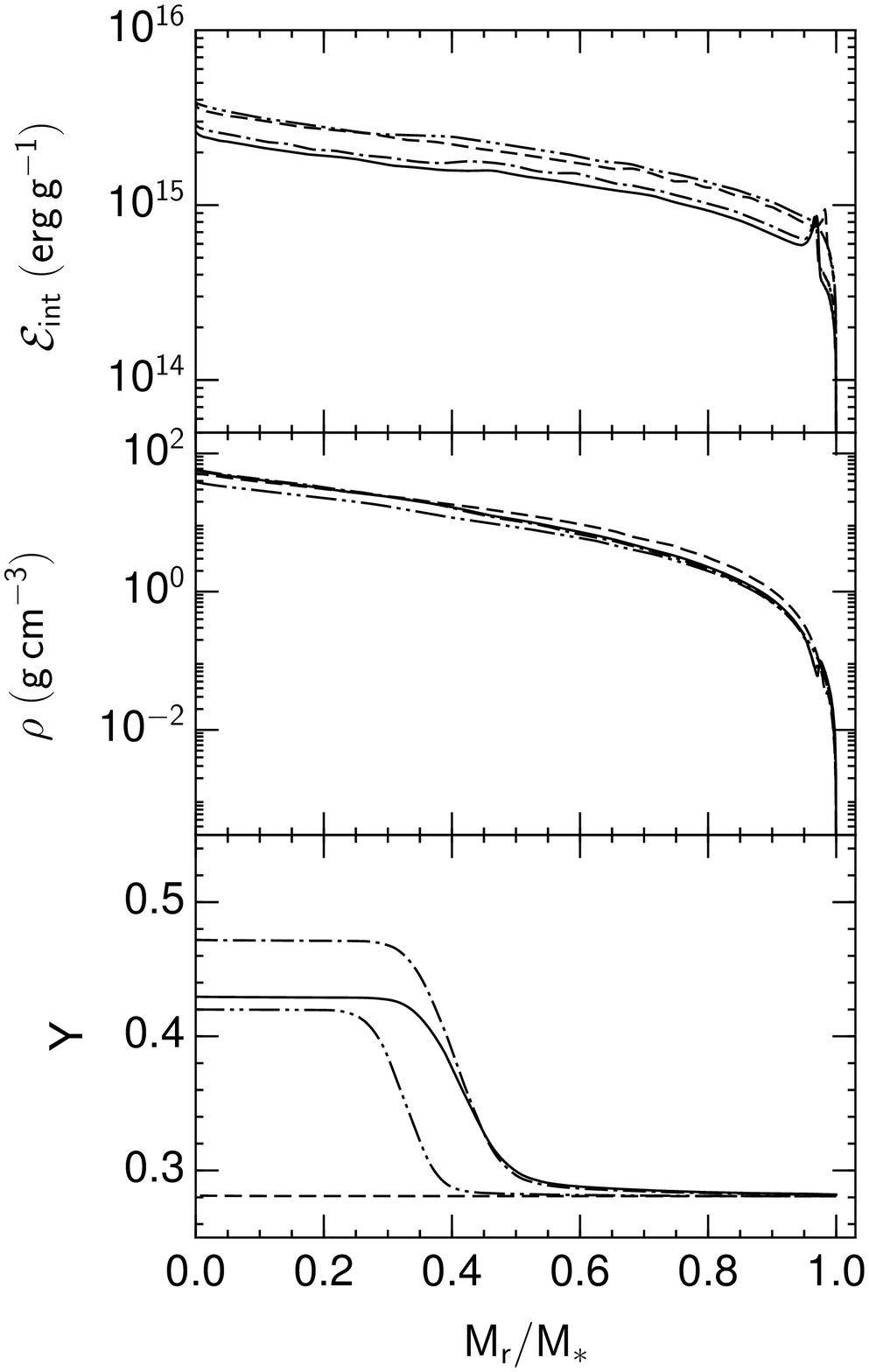}}
    \caption{Left-hand panel: Post-impact 1D angle-averaged profiles of specific internal energy $\mathcal{E}_{\mathrm{int}}$, density $\rho$, and He abundance $\mathrm{Y}$ as functions of fractional mass for four MS companion models at the end of SPH impact simulations of \citet[][see their table~1]{Liu2013c}, i.e., Model\_A, Model\_B, Model\_C, and Model\_D. Right-hand panel: similar to left-hand panel, but for the relaxed starting models in \textsc{MESA}.}
\label{Fig:3}
  \end{center}
\end{figure*}

\begin{figure*}
  \begin{center}
    {\includegraphics[width=0.48\textwidth, angle=360]{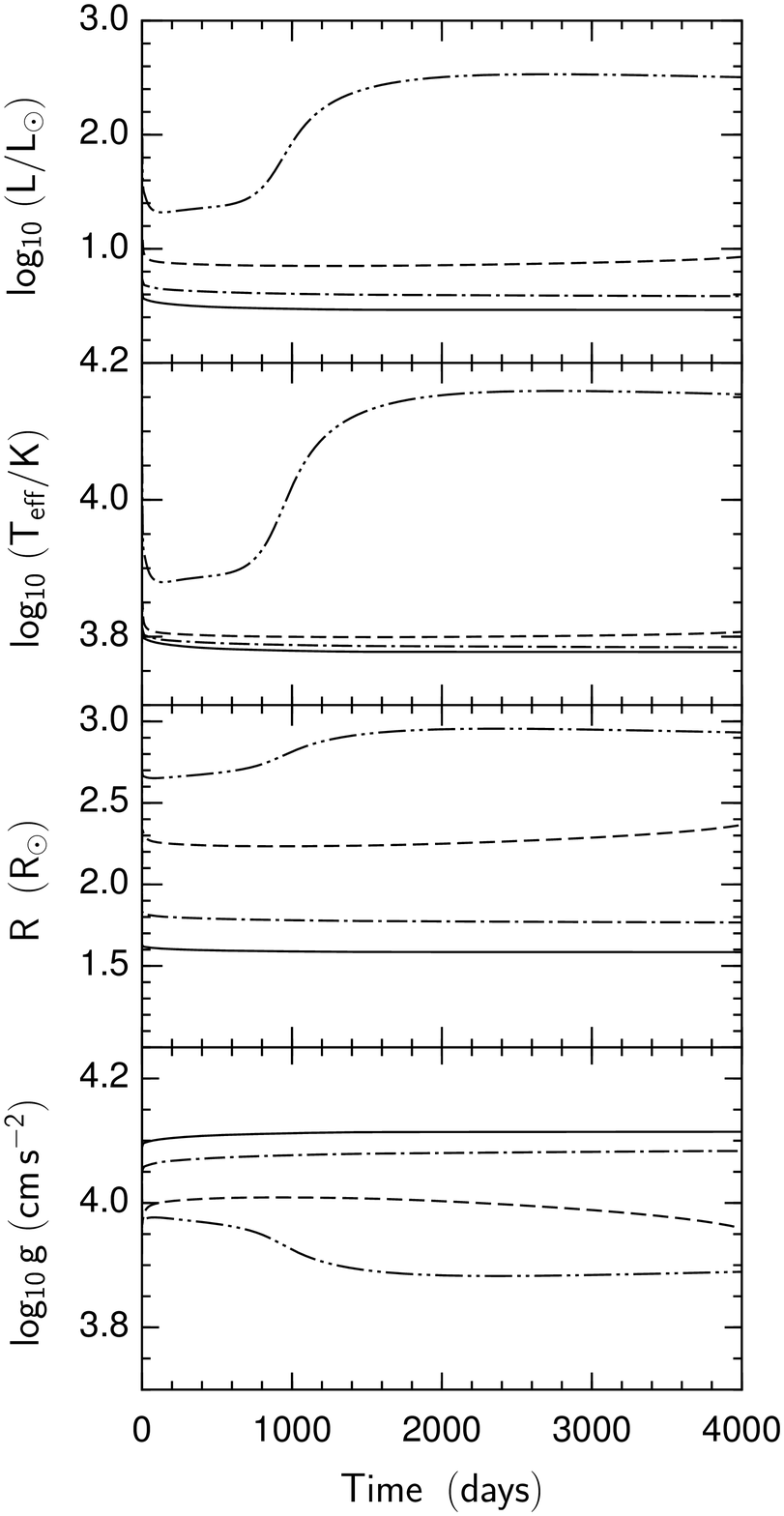}}
        \hspace{0.2in}
    {\includegraphics[width=0.48\textwidth, angle=360]{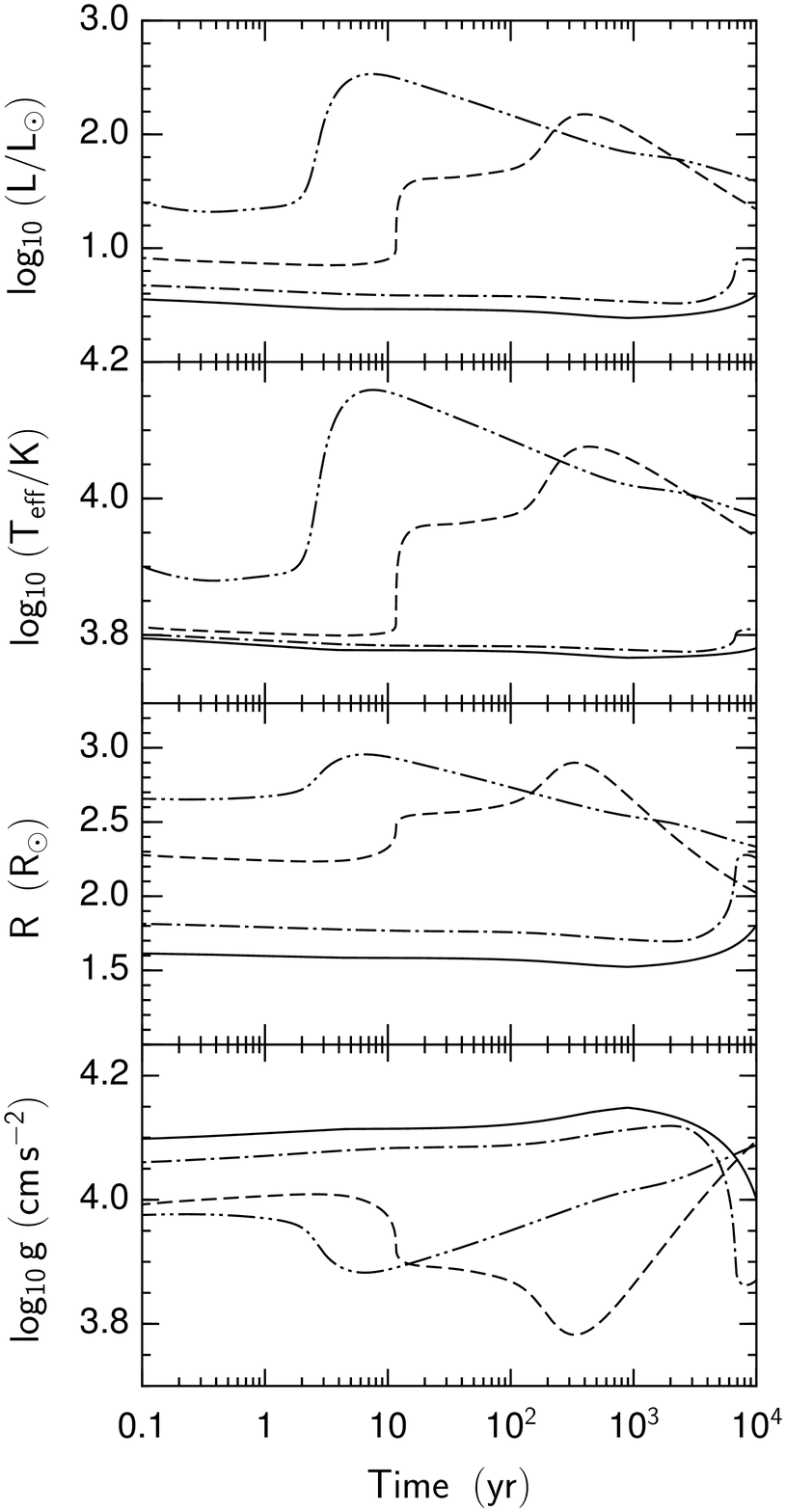}}
    \caption{Post-impact evolution of the photosphere luminosity $L$, effective temperature $T_{\mathrm{eff}}$, radius $R$, and surface gravity $g$ of surviving MS companion stars as functions of time. The solid, dash-dotted, dashed, and double-dotted curves correspond respectively to Model\_A, Model\_B, Model\_C, and Model\_D. The left-hand and right-hand panels are given with linear and logarithmic time axes, respectively.}
\label{Fig:4}
  \end{center}
\end{figure*}

In the present work, we map MS companion models obtained from 3D impact hydrodynamical simulations of \citet{Liu2013c} into the 1D stellar evolution code \textsc{MESA} \citep{Paxton2011, Paxton2018} to trace their long-term post-impact evolution. Although long-term post-impact evolution of surviving MS companions of normal SNe Ia within the SD progenitor scenario has been done by different previous studies \citep[e.g.,][]{Podsiadlowski2003, Pan2012b, Pan2013, Pan2014, Shappee2013, Bauer2019}, it has never been addressed in the context of SNe Iax. The aim of the present work is to explore the post-explosion observational features of surviving MS companions of SNe Iax by combining the outcome of 3D SPH impact simulations into the 1D stellar evolution code. It is expected to be helpful for searching for the surviving companions and therefore examining the valid of the SD binary systems as progenitors of SNe Iax. The method and companion models used in the present work are described in Section~\ref{sec:method}. Long-term evolution and appearance of four different MS companion models such as the luminosity and evolutionary tracks are presented in Section~\ref{sec:results} before we make discussions in Section~\ref{sec:discussion}. Finally, the important results and conclusions of this work are summarized in Section~\ref{sec:summary}.

\begin{figure*}
  \begin{center}
    {\includegraphics[width=0.48\textwidth, angle=360]{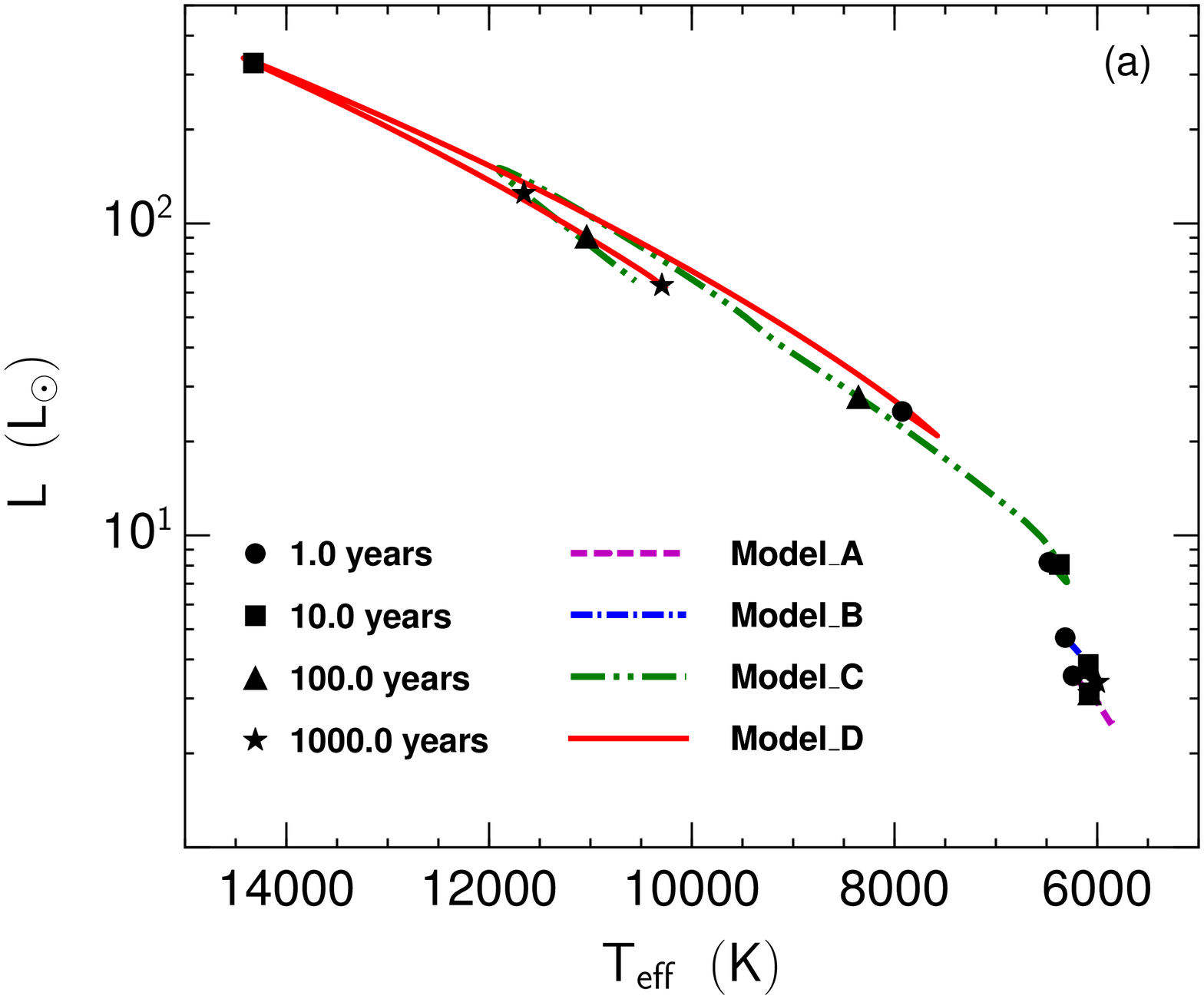}}
        \hspace{0.2in}
    {\includegraphics[width=0.48\textwidth, angle=360]{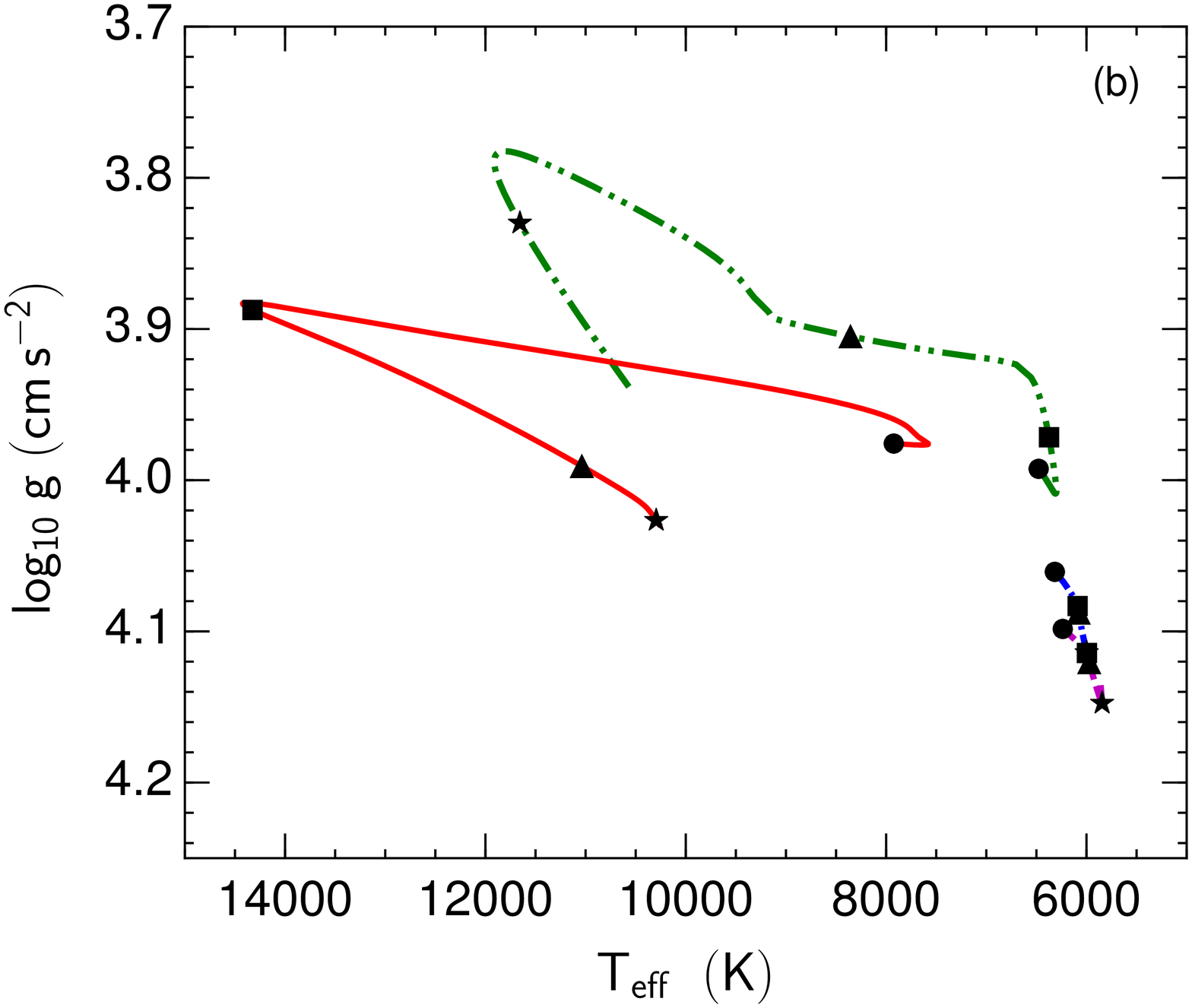}}
    \caption{Post-impact evolutionary tracks of Model\_A (dashed line), Model\_B (dash-dotted line), Model\_C (double-dotted line), and Model\_D (solid line). The filled circle, square, triangle, and star markers in the tracks present post-impact evolutionary phases of $1$, $10$, $100$, and $1000\,\mathrm{yr}$ after the SN impact, respectively. Panels (a) and (b) give the Hertzsprung-Russell (H-R) and surface gravity vs. temperature diagram, respectively.}
\label{Fig:5}
  \end{center}
\end{figure*}

\begin{figure*}
  \begin{center}
    {\includegraphics[width=0.48\textwidth, angle=360]{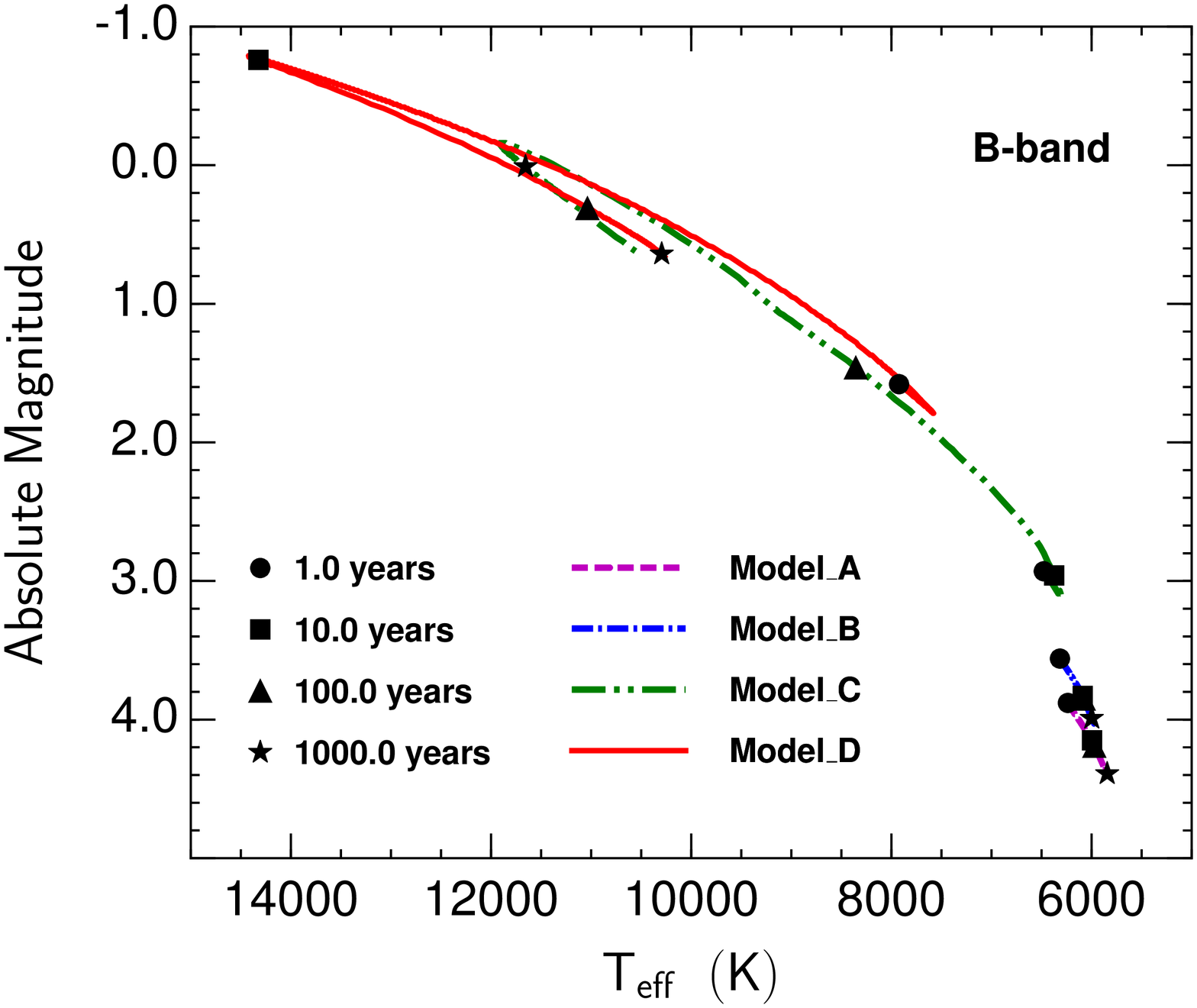}}
        \hspace{0.2in}
    {\includegraphics[width=0.48\textwidth, angle=360]{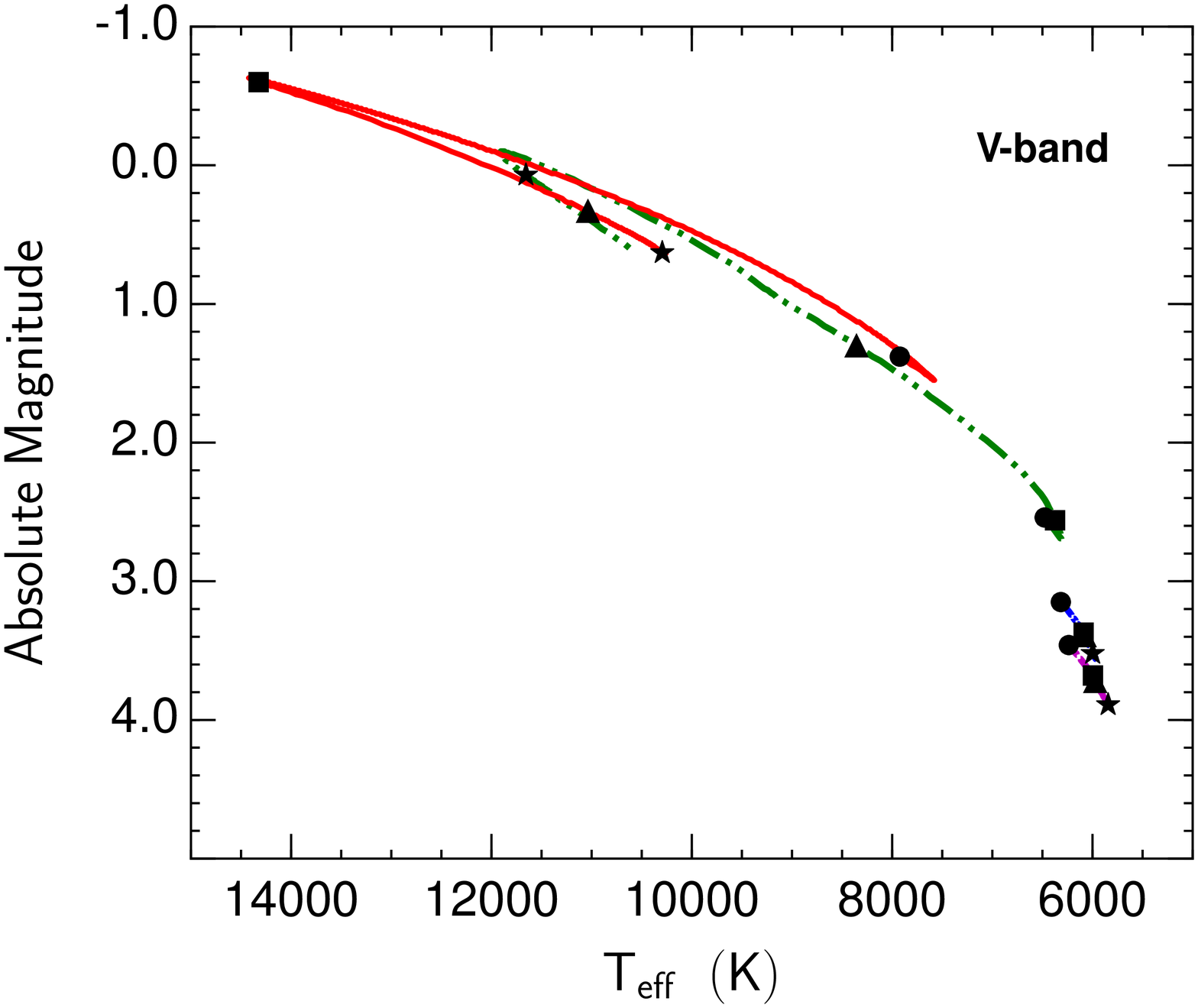}}
    \caption{Similar to Fig.~\ref{Fig:5}, but for $B$-band (left-hand panel) and $V$-band (right-hand panel) magnitude as functions of temperature. }
\label{Fig:6}
  \end{center}
\end{figure*}

\section{Methods and companion models}
\label{sec:method}

In this work, we assume that SNe Iax are generally produced from the WD~+~MS progenitor systems. To investigate long-term post-impact observational properties of surviving MS companion stars, we have performed 3D SPH simulations of SN~ejecta-companion interaction in \citet{Liu2013c}. Fig.~\ref{fig:remnant} shows density distribution of bound companion material in the orbital plane (i.e. $x-y$ plane) as a function of time from the 3D impact simulation of \citet{Liu2013c} for the Model\_A (see their table~1). At the beginning, the companion star is in a hydrostatic equilibrium state. The SN~ejecta-companion happens about a few minutes after the SN Iax explosion, removing some H-rich material from the right-hand side of companion surface. As the shock passes through the whole star, more H-rich material is removed from another side of the companion surface. At the end of simulation, the MS companion loses about $\lesssim 1\%$ of its mass due to the stripping and ablation by the SN ejecta, which depends on the progenitor model \citet[][see their table~1]{Liu2013c}. On the one hand, the MS companion star is significantly shocked and heated during the ejecta-companion interaction, leading to that the companion star puffs up dramatically at the end of simulations (about $t=5000\,\mathrm{s}$ after the explosion). On the other hand, the post-impact companion stars are not yet in hydrostatic and thermal equilibrium although it is being spherically symmetric at the end of 3D SPH simulations.

To provide post-impact observational properties of surviving companion stars for their identifications in historical SNRs, it is important to simulate and trace the detailed evolution of surviving companion stars for a long time up to the age of historical SNRs, i.e., a few 10--1000 yr. However, it is really difficult to run 3D hydrodynamical impact simulations for that long time because such simulations have a small time-step which is determined by the dynamical time-scale of the MS companion that is on the order of hundreds of seconds. Therefore, in the present work, 3D SPH post-impact companion models directly taken from 3D impact simulations of \citet[][]{Liu2013c} are mapped into the 1D stellar evolution code \textsc{MESA} \citep{Paxton2011, Paxton2018} to trace their subsequent long-term evolution \citep[see also][]{Pan2012b, Pan2013}.

The first step is to convert 3D post-impact companion models into angle-averaged 1D radial profiles. Therefore, we divide the final 3D companion models of SPH impact simulations (see the bottom right panel of Fig.~\ref{fig:remnant}) into 100--200 spherical shells. The parameters such as the internal energy ($\mathcal{E}_{\mathrm{int}}$), composition, density ($\rho$) of SPH particles in each shell are averaged to give a value for that shell. Fig.~\ref{Fig:2} presents an example of the comparison between 3D density distributions and the angle-averaged 1D radial profile of a surviving companion model, Model\_D. However, these 1D angle-averaged MS companion models cannot be used directly by \textsc{MESA} because they are still out of the hydrostatic equilibrium at the end of SPH impact simulations. Therefore, the second step is to import suitable starting models for the \textsc{MESA} based on the obtained 1D angle-averaged radial profiles of SPH models. In this work, by following the method described in \citet[][see their appendix B]{Paxton2018}, we directly used the relaxation routines provided in \textsc{MESA} to compute suitable starting models for \textsc{MESA}. These relaxation routines were developed to construct a starting model in hydrostatic equilibrium based on the specified profiles for entropy, composition, and angular momentum.  It has been shown that these relaxation routines can be successfully used for tracing long-term evolution of the outcome of a stellar merger from SPH simulations \citep[][see appendix B]{Paxton2018}. The comparison between pre-relaxation 1D density profile (green dashed line) and post-relaxation one (red solid line) for Model\_D is shown in Fig.~\ref{Fig:2}.

In this work, four different post-impact MS companion models (i.e. the so-called Model\_A, Model\_B, Model\_C, and Model\_D) are directly taken from 3D SPH impact simulations of \citet[][see their table~1]{Liu2013c}. The detailed properties of these four models at the moment of SN explosion (i.e. before ejecta-companion interaction) can be found from \citet[][see their table~1 and Fig.~8]{Liu2012} and \citet[][]{Liu2013c}. The spherical averaged 1D profiles of the composition, $\mathcal{E}_{\mathrm{int}}$, and $\rho$ of these post-impact models are put into the relaxation routines of \textsc{MESA} to compute the starting models for their subsequent long-term post-impact calculations. Fig.~\ref{Fig:3} shows the angle-averaged 1D profiles (left-hand panel) of $\mathcal{E}_{\mathrm{int}}$, helium abundance $Y$, and $\rho$ of four post-impact MS models at the end of SPH simulations of \citet{Liu2013c}. The corresponding profiles of the post-relaxed models in \textsc{MESA} are also given in right-hand panel of Fig.~\ref{Fig:3}. Note that a zero angular momentum is set when constructing the starting model for \textsc{MESA} because the orbital motion of the binary system and rotation of a star were not included into SPH impact simulations of \citet{Liu2013c}. The effect of orbital motion of the binary and rotation of a star on the results will be investigated by a forthcoming study.

\section{Post-impact evolution of surviving companions}
\label{sec:results}

By using the method described in Section~\ref{sec:method}, post-impact 1D averaged profiles in left-hand panel of Fig.~\ref{Fig:3} have been put into the relaxation routines of \textsc{MESA} to compute the relaxed starting models (right-hand panel of Fig.~\ref{Fig:3}). We then trace long-term evolution of these models to make predictions on the post-impact observational properties such as temperature, luminosity, radius, and surface abundances. This is will be helpful for current and/or future observations to identify surviving companion stars of SNe Iax and therefore favor (or disfavour) the SD progenitor origin of SNe Iax.

\subsection{Typical time evolution}

Figure~\ref{Fig:4} shows time evolution of post-impact photosphere luminosity $L$, effective temperature $T_{\mathrm{eff}}$, radius $R$, and surface gravity $g$ as functions of time for four different MS companion models. As mentioned above, the companion stars are significantly shocked and heated by the SN impact during the interaction, leading to that they expand dramatically after the SN explosion due to the release of deposited energy. Depending on not only the different MS companion models, but also the total amount of injected energy and the depth of energy deposition during the interaction, the post-impact MS companion stars continue to expand on a time-scale of $10$--$10^{4}$ yr (from Model\_D to Model\_A) before they start to contract\footnote{This expansion timescale is determined by the local thermal timescale of the surviving companion star \citep{Henyey1969}.}. As a result, the companion stars become significantly more luminous (i.e. about $5$--$500\,L_{\sun}$ from Model\_A to Model\_D, see right-hand panel of Fig.~\ref{Fig:4}). As the deposited energy radiated away, the companion starts (e.g. about 10 yr after explosion) to contract by releasing the gravitational energy, leading to the decease of its radius and luminosity (Fig.~\ref{Fig:4}).

\begin{figure}
  \begin{center}
    {\includegraphics[width=0.49\textwidth, angle=360]{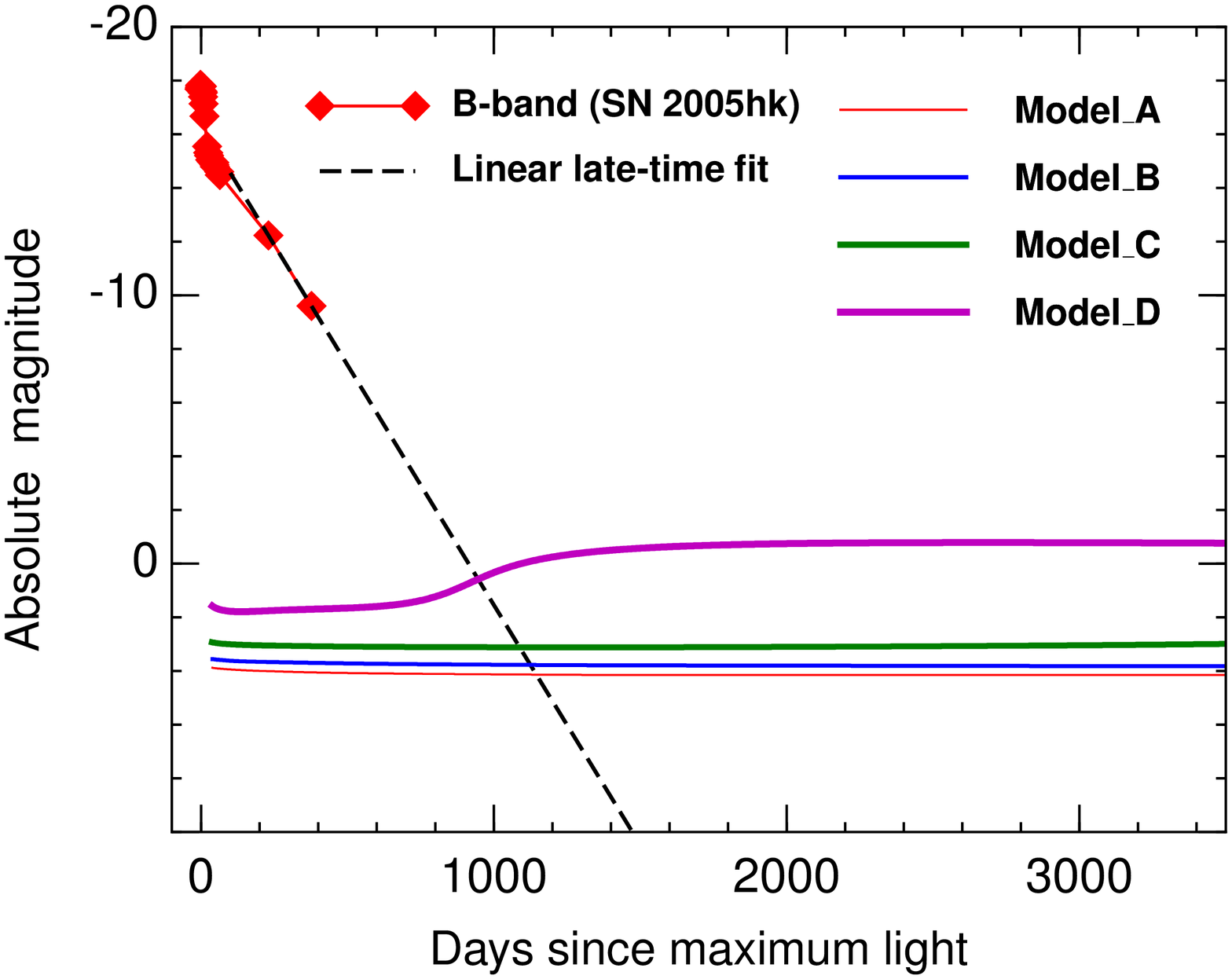}}
    \caption{Comparison between the post-impact light curves of surviving MS companion stars of SNe Iax and late-time light curve of SN~2005hk (red diamonds, \citealt{Phillips2007, Sahu2008}) with linear late-time decline fit (dashed line). The predictions of different surviving MS companion star models are given in solid lines with a different thickness.}
\label{Fig:7}
  \end{center}
\end{figure}

Basic post-impact evolution of a MS companion star in this work is similar to that in \citet{Pan2012b}. However, their post-impact companions generally become more luminous than our models. This is because that \citet{Pan2012b} focused on normal SNe Ia rather than SNe Iax, and the `W7 model' of \citet{Nomoto1984} was therefore used to represent the SN explosion. Compared with the N5def model ($1.34\times10^{50}\,\mathrm{erg}$) used in this work , a higher kinetic energy ($1.23\times10^{51}\,\mathrm{erg}$) of the W7 model leads to a higher amount of energy deposition during the interaction and thus a more luminous post-impact object before contracting. By using the method of \citet{Pan2012b}, we have calculated the energy deposition by tracing the amount of increased binding energy of the star after the SN impact, which is about $1.6$--$2.0\times10^{48}\,\mathrm{erg}$ in our models. Based on the ratio of binary separation to companion radius at the moment of SN explosion (i.e. $a_{\mathrm{f}}/R_{\mathrm{2,f}}$, see \citealt{Liu2013c}), we can estimate that the total incident energy in our models is about $4.5$--$6.1\times10^{48}\,\mathrm{erg}$, suggesting that about $26$--$40\%$ of the incident energy is injected into MS companion stars during the interaction.

\subsection{Evolutionary tracks}

Figure~\ref{Fig:5} shows post-impact evolutionary tracks of different MS companion models in the Hertzsprung-Russell (H-R) and effective temperature-surface gravity ($T_{\mathrm{eff}}$--$g$) diagram. Under the assumption that the companion photosphere emits a blackbody radiation spectrum,  the effective temperature ($T_{\rm{eff}}$) and photospheric radius ($R$) of a post-impact companion star model in \textsc{MESA} have been used to convert the bolometric luminosity to broad band magnitudes \citep[see also][]{Pan2013,Liu2015}: 
 
 \begin{equation}
    \label{eq:}
m_{S_{\lambda}} = -2.5\,\rm{log_{10}}\,\left [\frac{\int S_{\lambda} (\pi B_{\lambda})d\lambda}{\int (\textit{f}^{\,0}_{\nu}\,c/\lambda^{2})S_{\lambda}d\lambda}\ \left(\frac{R}{d}  \right)^{2} \right ]  
  \end{equation}

where $S_{\lambda}$ is the sensitivity function of a given filter at wavelength $\lambda$, $B_{\lambda}$ is the Planck function, $d$ is the distance of the star, and $f^{0}_{\nu}=3.631\times10^{-20}\,\rm{erg\,cm^{-2}\,s^{-1}\,Hz^{-1}}$ is 
the zero-point value in the AB magnitude system. Specifically for this work, the different filters of the `Bussell-UBVRI system' and their corresponding sensitivity functions are considered. Fig.~\ref{Fig:6} gives the B-band and V-band magnitude as functions of temperature.

\section{Discussions}
\label{sec:discussion}

\subsection{Comparison with the observations}
\label{sec:survivor}

In the SD progenitor scenario, the companion star is expected to survive from the SN explosion. Therefore, whether or not the existence of the shocked surviving companion star could be detected at late-time observations of SNe Iax provides a promising way to examine the valid of the studied progenitor model and explosion mechanism, or to place constraints on them. In Fig.~\ref{Fig:7}, the time evolution of predicted $B$-band magnitudes of four different surviving MS companion models is compared with the late-time light curve of a prototypical Iax event, SN~2005hk \citep[][]{Phillips2007, Sahu2008}. As it is shown, the post-impact luminosity of a surviving companion can exceed that of SN ejecta in $B$-band about 1000 days after the maximum light. This suggests that the surviving companion stars would begin to dominant late-time light curve of SNe Iax about 3 yr after the explosion. However, the late-time light curve of SN~2005hk in Fig.~\ref{Fig:7} is simply given with a linear decline fit. As discussed by \citet[][see their Section~3.2.2]{Shappee2013}, the real late-time light curve could be flatter due to some effects such as the decay of $^{57}\mathrm{Co}$ \citep[][]{Roepke2012}, leading to that the surviving companion stars would be observable at later times.

\subsection{Implications of the bound WD remnant}

In this work, SNe Iax are assumed to be generally generated from incomplete pure deflagration explosions of near-Chandrasekhar-mass WDs in SD progenitor systems. Based on current hydrodynmic modelling of this explosion model, pure deflagration explosion does not burn the complete Chandrasekhar-mass WD, leaving behind a bound WD remnant that is significantly heated by SN explosion and enriched with iron group elements (IGE) material and radioactive isotopes such as $^{56}\rm{Ni}$ due to fallback of SN explosion ashes \citep[e.g.][]{Jordan2012a, Kromer2013}. Therefore, it is expected that this bound WD remnant will dramatically puff up after the explosion, contributing the late-time light curves of SNe Iax and possibly becoming detectable when SN itself has faded.

In particular, the N5def pure deflagration model of \citet{Kromer2013} and \citet{Fink2014} has been used in this work. In this model, the SN explosion leaves behind a $1.0\,M_{\sun}$ bound WD remnant. However, given the strong expansion of the SN ejecta, it is difficult to spatially resolve this bound WD remnant until late-time \citep{Fink2014}. By constructing a simplified initial condition to mimic the results of the bound WD remnant of \citet{Fink2014}, \citet{Shen2017} have explored the post-explosion evolution of the bound WD remnant of SNe Iax \citep[see also][]{Zhang2019} with the stellar evolution code \textsc{MESA} with considering the effect of delayed decays of captured radioactive isotopes such as $^{56}\mathrm{Ni}$  and $^{56}\mathrm{Co}$. They found that such delayed decays can drive a persistent wind from the WD surface, and the predicted observational properties of the WD remnant match the late-time luminosities of observed SNe Iax such as SN~2005hk, SN~2008A, and SN~2008ha, providing evidence that SNe Iax are resulted from the incomplete deflagration of a near-Chandrasekhar mass WD \citep{Shen2017}. However, the simplified initial conditions in their modelling could lead to the uncertainties of their results. To better estimate observable signatures of the bound WD remnant, future works with accurate initial conditions are still encouraged.

\subsection{More features of survivors}

In the WD~+~MS progenitor systems, the MS companions have an orbital velocity of  $\upsilon_{\mathrm{orb}}\approx70$--$240\,\mathrm{km\,s^{-1}}$ and a surface rotational velocity of $\upsilon_{\mathrm{rot}}\approx40$--$170\,\mathrm{km\,s^{-1}}$ at the moment of the explosion \citep[e.g.][]{Han2004, Liu2018}\footnote{Here, the MS donor star is assumed to be tidally locked to its orbital motion.} In addition, the MS companion receives an impact velocity of typically about $\upsilon_{\mathrm{kick}}\lesssim 30\,\mathrm{km\,s^{-1}}$ during the SN ejecta-companion interaction \citep{Liu2013c}. Therefore, it is expected that the surviving MS companions would have a spatial velocity of $\upsilon_{\mathrm{spatial}}$=$\sqrt{\upsilon_{\mathrm{orb}}^{2}+\upsilon_{\mathrm{kick}}^{2}}\lesssim 76$--$241\,\mathrm{km\,s^{-1}}$ if the WD~+~MS binary system can be destroyed by the explosion. Besides the overluminous signatures presented above, a high spatial velocity provides another feature for identifying the surviving MS companion star of SNe Iax. 

However, whether or not the WD~+~MS binary system would be destroyed after the explosion strongly depends on the kick velocity received by the WD remnant and the amount of unbound WD masses. But unfortunately, the kick velocity of the bound WD remnant is still largely uncertain based on current simulations \citep[e.g.][]{Jordan2012a, Kromer2013, Fink2014}. Given this uncertainty, instead of leading to the disruption of the binary system, and therefore generating a fast-moving surviving companion star and bound WD remnant after the explosion, it is also possible that the new binary system composed of two strongly heated components by ejecta-companion interaction and explosion itself would survive. Furthermore, this surviving binary system might evolve and merge into a single object with a rapid rotation velocity, or experience a common envelope phase \citep[][see their Sec.~5.4]{Liu2013c}. An accurate prediction on the post-explosion fate of a binary progenitor system can only be given until future explosion modelling of SNe Iax could provide an accurate estimate on the kick velocity and detailed structures of the bound WD remnant.

\subsection{Different progenitor systems}

Although the present work concentrates on the MS donor models, the non-degenerate companion star could also be a red giant and/or a helium star in the SD progenitor systems. There is observational evidence indicating that the helium star donor binary systems are likely to be the potential progenitors of SNe Iax \citep[][]{Foley2013}. For instance, a blue bright source that was detected in pre-explosion images of an SN Iax, SN~2012Z, has been suggested to be the possible helium companion star of its progenitor system \citep{McCully2014}. To comprehensively provide predictions on the observational features of surviving companion stars of SNe Iax, it would be important to also explore the long-term evolution and appearance of surviving helium companion stars. Very recently, 3D hydrodynamical simulations of the impact of SN Iax ejecta on a helium companion star have been studied in detail by \citet{Zeng2020}. Post-impact observational signatures of surviving helium companion stars will be presented in our forthcoming paper (Zeng et al. in preparation).

\subsection{Model uncertainties}

As mentioned above, a specific N5def explosion model was used in 3D hydrodynamical simulations of ejecta-companion interaction by \citet{Liu2013b}. Although the N5def model has been found to be able to reproduce the characteristic observational features of a prototypical SN Iax (i.e. SN 2005hk, $M_\mathrm{V,peak}= -18.1\,\mathrm{mag}$), the brighter (e.g. SN~2012Z, $M_\mathrm{V,peak}= -18.5\,\mathrm{mag}$, see \citealt{McCully2014, Stritzinger2015}) and the fainter (e.g. SN~2008ha, $M_\mathrm{V,peak}= -14.2\,\mathrm{mag}$, see \citealt{Foley2009}) SNe Iax have also been observed. To explain observables of these brighter and fainter events, an explosion model with a higher and/or lower kinetic energy would be needed \citep[e.g.][]{Kromer2015}. As a result, the total amount and depth of the energy deposition of surviving companion stars by the interaction could be different, causing that they become more (or less) luminous than the models of this work in their long-term post-impact evolution. In addition, different explosion mechanism of a near-Chandrasekhar mass WD such as the pulsational delayed detonation explosion (PDD; \citealt{Hoeflich1995}) has also been proposed for SNe Iax \citep{Stritzinger2015}. The different explosion mechanism is also expected to influence the ejecta-companion interaction and thus the post-impact properties of surviving companion stars, which needs to be investigated in detail by future studies.

\section{Conclusion and Summary}
\label{sec:summary}

In this work, assuming that SNe Iax are generated from the incomplete deflagration explosions of near-Chandrasekhar mass WDs in WD~+~MS binary progenitor systems, we have explored long-term post-impact evolution and appearance of surviving MS companion stars by mapping the output model of 3D hydrodynamical simulations of ejecta-companion interaction by \citet[][]{Liu2013b} into the 1D stellar evolution code \textsc{MESA} \citep[][]{Paxton2011, Paxton2013, Paxton2015, Paxton2018}. Our main results and conclusions can be summarized as follows: 

\begin{itemize}\itemsep5pt 
\item  Depending on different MS companion models, the total amount of injected energy and the depth of energy deposition by SN impact are different. In our models, it is found that about $26$--$40\%$ of the total incident energy is injected into MS companion stars during the ejecta-companion interaction.

\item  As a result, the surviving MS companion stars continue to expand on a time-scale of $10$--$10^{4}$ yr before they start to contract, leading to that they become significantly more luminous (i.e. about $5$--$500\,L_{\sun}$ from Model\_A to Model\_D, see right-hand panel of Fig.~\ref{Fig:4}) in their post-impact evolution.

\item  Comparing our results with the late-time light curve of a prototypical SN Iax, SN~2005hk, it is found that the predicted luminosities of surviving companion stars start to exceed that of SN ejecta in $B$-band about 1000 days after the explosion. Under an assumption that SNe Iax are generally generated from the incomplete deflagration explosions of near-Chandrasekhar-mass WDs in WD~+~MS progenitor systems, it is therefore suggested that their surviving companion stars would begin to be visible when SN itself has faded about 3 yr after the explosion.

\end{itemize}

\section*{Acknowledgements}

We would like to thank the anonymous referee for reviewing our manuscript. This work is supported by the National Natural Science Foundation of China (NSFC, No. 11873016), the Chinese Academy of Sciences and the Natural Science Foundation of Yunnan Province of China (No. 202001AW070007). The authors gratefully acknowledge the ``PHOENIX Supercomputing Platform'' jointly operated by the Binary Population Synthesis Group and the Stellar Astrophysics Group at Yunnan Observatories, Chinese Academy of Sciences.

\section*{DATA AVAILABILITY}
The data underlying this article will be shared on reasonable request to the corresponding author.

\bibliographystyle{mnras}

\bibliography{ref}

\end{document}